\documentstyle[12pt]{article}

\begin{document}
\thispagestyle{empty}

\begin{center}

\null
\vskip-1truecm
\rightline{IC/IR/96/17}
\vskip1truecm
{International Atomic Energy Agency\\
and\\
United Nations Educational Scientific and Cultural Organization\\
\medskip
INTERNATIONAL CENTRE FOR THEORETICAL PHYSICS\\}
\vskip2truecm
{\bf
CONFINEMENT IN A DOUBLE BARRIER STRUCTURE IN THE PRESENCE OF
AN ELECTRIC FIELD\\ }
\vskip2truecm
{ S.M.A.Nimour, N.Zekri and R.Ouasti\\
International Centre for Theoretical Physics, Trieste, Italy\\
and\\
Laboratoire d'Etude Physique des Materiaux, Departement de Physique
U.S.T. Oran, B.P. 1505 El M'Naouar  Oran  Algerie\footnote{\normalsize
Permanent Address }\\}
\end{center}
\vskip1truecm
\centerline{ABSTRACT}
\baselineskip=24pt
\bigskip

The effect of electric field on the electron resonant tunnelling into a
double barrier structure is studied. We show for particular field strengths an
increase of the tunnelling time which leads us to explain the Stark-ladder
localization and to discuss Bloch oscillations and the quenching of luminescence
in multiple quantum wells.
\vskip.5truecm
\begin{tabbing}
111 \= 2222222222222 \= \kill

\> \underline{\bf Keywords}: \>Resonant Tunnelling, Double Barrier,
                                         Stark-Ladder Localization,\\
\>                           \> Bloch Oscillation, Quantum Wells
\end{tabbing}
\vskip1.5truecm
\begin{center}
MIRAMARE -- TRIESTE\\
\medskip
July 1996
\end{center}

\newpage

\baselineskip=18pt

\section{ INTRODUCTION}
    Since its prediction  \cite{kn:james,kn:wannier} the Wannier-Stark
ladder localization has been difficult to observe experimentally \cite{kn:mendez1}.
This effect takes place when the electron wave-vector becomes of the order of
the reciprocal lattice parameter leading to Bragg reflections. During the last
decade several papers examined this effect theoretically \cite{kn:tsu} and
experimental evidence has been provided by photoluminescence measurements
\cite{kn:mendez2}. The effect has now a wide application to the optical properties
of quantum wells \cite{kn:harwit} and photodetectors based on the effect
have been developed recently \cite{kn:levine}. Moreover, useful resonant
tunnelling current-voltage peaks can be made by engineering the well width and
the barrier height \cite{kn:bauer}.
   Resonant tunnelling is a quantum phenomenon in which the electron
wave-function interference becomes constructive. In the case of a single quantum
well it corresponds to a quasi-bound state while in a symmetric double barrier
structure it shows a transmission coefficient peak characterized by the energy
position $E$ and the width $\Delta E$. The analytical resonant tunnelling condition
determines these parameters for a given double barrier structure \cite{kn:yamamoto}
and shows a strong dependence on the width of the well. The applied electric field
provides a way to increase the Fermi energy of the conduction band electrons
which shifts the resonant peaks and minibands.
   The resonant tunnelling width can be related to the corresponding lifetime by
\cite{kn:harada}:
\begin{equation}
\tau = \hbar / \Delta E
\end{equation}
We can interpret the motion of the carriers semi-classically by using the group
velocity of an electron wave-packet. We also expect that, in an electric
field, the electron wave-packet will be accelerated leading to a decrease of
its lifetime. The resonant tunnelling width will therefore broaden in an
electric field. This has been shown by the recent work of Gonzalez et al. \cite{kn:gonzalez}.
However their calculations were made for high fields (from 200 to 400 kV/cm)
for which the electric field can lead to the localization of the electron states
which confines them in a small region of the sample. In particular, it has been
shown recently for one-dimensional $\delta$-peak potentials, that localization appears at
lower fields and should disappear when the field strength is increased \cite{kn:zekri}.
Therefore, we expect  different behaviours of the tunnelling time in such a case
in the quantum well and double barrier systems.
        In this paper we discuss the behaviour of the electron tunnelling time
by calculating the transmission coefficient in the case of the Stark-ladder
localization for a double barrier system. An extensive study of electron
transmission in an electric field leads us to give a new interpretation of the
origin of the field induced localization in periodic systems and the
transmission oscillations observed recently \cite{kn:zekri}. We expect a decrease of the
tunnelling time to yield a deceleration of the electron wave-packet. Similar
considerations are involved in the quenching of the luminescence in the
$GaAs-Ga_{1-x}Al_xAs$ multiple quantum wells \cite{kn:mendez3}.

\section{ THE MODEL}
        We consider a single electron in a rectangular double barrier potential
in a constant electric field. The corresponding Schrodinger equation can be read:
\begin{equation}
{\hbar^2\over 2m^{*} }{\partial^2\over\partial x^2}\Psi + V(x)\Psi - \mid e\mid Fx
= E\Psi
\end{equation}
where the potential $V(x)$ is defined for a well and a barrier of width Lw and Lb
respectively  by:
\begin{equation}
V(x) = \left\{
  \begin{array}{l l}
V_0  &: 0 < x < L_b \quad {\rm or } \quad L_b + L_w < x < L_w + 2L_b \\
&  \\
0    &: {\rm otherwise}
  \end{array}
\right.
\end{equation}
The solutions of Eq. (2)  are a combination of independent Airy functions. We use
the transfer matrix method to determine the transmission coefficient through
the structure (the detailed method is given by Cota et al. \cite{kn:tsu}). The resonant
energy width is identified with  the half maximun width by assuming that the
transmission spectrum is Lorentzian near  resonance.

\section{ DISCUSSION OF THE RESULTS}
        We are interested in the electron confinement in the growth direction
for a double barrier structure in the presence of an electric field. We therefore
calculate the transmission spectrum for field strengths between 0 and 200 kV/cm
since above 200 kV/cm the tunnelling time decreases \cite{kn:gonzalez} leading to the acceleration
of the electron wave-packet. Furthermore, Zekri et al. \cite{kn:zekri} show that Stark
localization can be observed clearly for weak fields and tends to disappear for
higher ones. These authors also show (using a chain of $\delta$-peak potentials)
that a breakdown of the minibands takes place and a set of localized states
appears. This localization does not depend on the
potential model and can also be observed in a chain of rectangular potentials as
shown in Fig.1 where the transmission corresponding to only a part of the
minibands disappears because of the electric field.

        In the following calculations the barrier and the well effective mass
are taken to be  0.067 and 0.108  respectively with widths 20 and 50 $\AA$
respectively while the potential height of both barriers is 500 meV. With the
chosen parameters we have two resonant transmission peaks and can compare the
effect of the field on each peak.

        In Fig.2a the transmission coefficient for a double barrier structure
is shown for different values of the field F=0, 50 and 100 kV/cm. The main
effect of the field is to shift the resonant peaks which can be understood in
terms of the increase of the kinetic energy of the conduction electrons since,
at the resonant energy, the electrons move freely through the barriers. Indeed
this is clearly shown in Fig.2b where the energy shift increases linearly with
the field for both peaks and its slope has the magnitude of |e|L (L being the
width the well). Therefore the new shifted resonant energy for this system is:
\begin{equation}
E_f = E_0  + \mid e\mid FL
\end{equation}
where E is the zero field resonant energy. However, this behaviour is not in
agreement with the results of Gonzalez et al. \cite{kn:gonzalez} and Bastard et al.
\cite{kn:bastard} where the shift behaves quadratically with the field.
In fact they considered a single quantum well while the structure studied here
is a double barrier. This difference between the two structures is due to a
change of the multiple reflection effect which is strong in the single quantum
well case at the bound state because the electron wave function becomes
evanescent in the infinite width of the potential barrier while for our
structure this reflection disappears at the resonant tunnelling energy and the
electron does not "see" the finite width of the potential barrier. However, an
infinite barrier width is unrealisable and in the practical realization of the
single quantum well, the potential barrier width is large enough to ensure that
the outgoing electron wavefunction becomes very small.

        The other main effect shown in Fig.2a is that the first resonant peak
becomes narrower with increasing field while the second one becomes broader.
This is clearly shown in Fig.3 where the full width at half maximum is plotted
as a function of the field strength for both resonant peaks and a fast drop is
observed for the first peak while the increase for the second one is to be linear.
From Eq. (1) the decrease of the width of the first peak means an increase
of the tunnelling time and leads to the confinement of the electron wave-packet
in the potential well. Also shown in Fig.2a are non-unity resonant peaks when
the field is applied. This decrease of the resonant transmission peaks may be
responsible for the breakdown \cite{kn:zekri} of the minibands in a superlattice
built from a double barrier since near this resonance the reflection vanishes
and the transmission coefficient for the superlattice becomes a high power of
the corresponding double barrier value.

    The decrease of the width of the first resonance leads to Bloch oscillations
in the energy spectrum. We recall that the minibands of a superlattice are
produced by the coupled resonant peaks of the multiple potential wells
\cite{kn:kouwenhoven}. If the number of potential wells increases, these
resonant energies become closer and the corresponding transmission spectrum
overlap to produce a continuum. When the electric field is applied it decreases
the overlap which leads to an oscillation of the transmission coefficient as
is clearly shown in the energy spectrum given in \cite{kn:zekri}.

   The field-induced localization disappears for higher fields \cite{kn:zekri}.
Such a delocalization arises from the existence of a critical field at which
the width of the first peak reaches a minimum and subsequently increases. This
minimum is not observed in Fig.3 because the first peak is also shifted by the
field and disappears before reaching the critical field. By using a different
structure, Agullo-Rueda et al. \cite{kn:agullo} have interpreted this Stark-localization
effect in terms of the interaction of the localized and the extended states.
However, this is not the case in our superlattice structure which has no
localized states in the absence of the field. The Stark localization observed
at the first resonance is not seen at the second one because the electron
wavelength at this energy becomes different from the distance between the two
barriers and then there is no Bragg reflection.

        The increase of the tunnelling time observed in Fig.3 may also be
responsible for the quenching of the luminescence observed by Mendez et al. \cite{kn:mendez3}.
This phenomena has been interpreted by Bastard et al. \cite{kn:bastard} as the effect of
the increase of the distance between the conduction electron and the hole
which leads to a decrease of the recombination between them but from their
results this is not sufficient to explain this experimental observation.
However, Buttiker and Landauer \cite{kn:buttiker} argued that the electron interacts with a
radiation of frequency $\omega$ if its tunnelling time $\tau$ satisfies
$\omega\tau \sim 1$.
Therefore the field-induced high tunnelling time observed at the first energy
resonance can lead to the emission of a low frequency radiation which may not
be observed in the luminescence experiments.

\section{ CONCLUSIONS}
        In this paper we have studied numerically the effect of the electric
field on a double barrier structure on the transmission coefficient. It is
found that, for particular field strengths, confinement appears due to the
decrease of the resonant tunnelling width and the transmission peak value.
These effects allow us to explain the Stark ladder localization and Bloch
oscillations recently calculated for a one-dimensional chain in an electric
field \cite{kn:zekri}. The localization can also explain (via the tunnelling time) the
quenching of the luminescence observed by Mendez et al. \cite{kn:mendez3} . However, a
quantitative study of this effect and an analytical determination of the
field values localizing the system are needed to fully interpret the Stark
ladder effects. Such a study will allow us to choose the parameters of the
structure to adjust the critical field at which the resonant energy width
is a minimum. The search for this minimum width is important since, if it is
sufficiently small, this can lead to an apparent violation of the Heisenberg
uncertainty principle. Moreover, the study of this effect in more realistic
structures, e.g. non-abrupt heterojunctions, is required to  allow us to make
quantitative comparisons with experiments.

\section{Acknowledgement}
        One of the authors (N.Z.) would like to thank the International Center for
Theoretical Physics  for hospitaly during his visit to the centre where part of
this work has been done to Professor P.N.Butcher for reading the manuscript. He
is very grateful to the Arab Fund for financial support during his visit to
the centre.

\newpage

\centerline{\bf Figure Captions}
\bigskip
\begin{description}
\item{Figure 1}
The transmission spectrum for a superlattice of 50 barriers in
the absence of electric field (solid curve) and for a field strength of
10 kV/cm (dotted curve).\\

\item{Figure 2.a}
The transmission spectrum for a double barrier for field
strengths of 0 (solid curve), 10 kV/cm (dotted curve) and 50 kV/cm
(long dashed curve). \\

\item{Figure 2.b}
The energy shift as a function of the field for both the
first and the second peak. The dotted line is used as a guide for the eyes.\\

\item{Figure 3}
The full width at half maximun as a function of the field for
the first peak (open losenges) and the second peak (+). The scale of the second
resonance widths should be multiplied by 10.

\end{description}

\newpage

\setlength{\unitlength}{0.240900pt}
\ifx\plotpoint\undefined\newsavebox{\plotpoint}\fi
\sbox{\plotpoint}{\rule[-0.200pt]{0.400pt}{0.400pt}}%


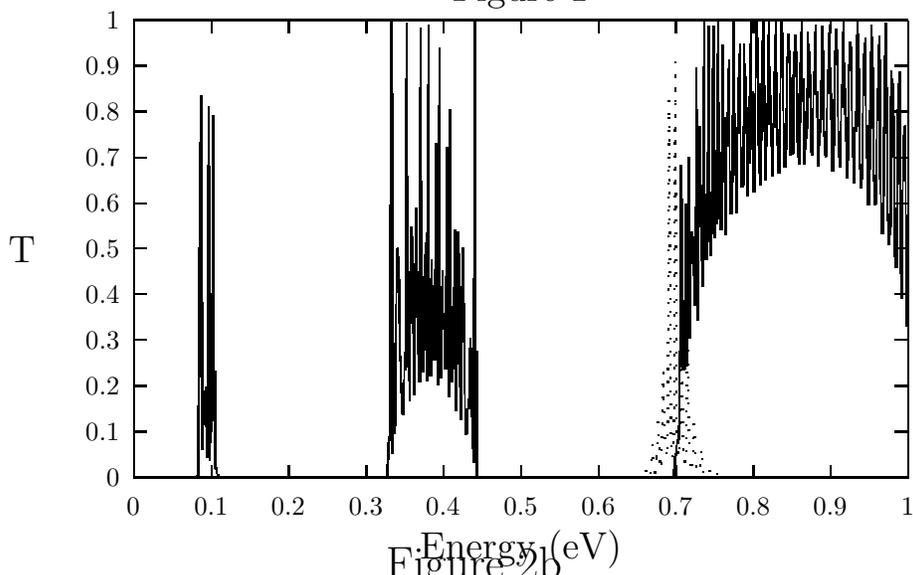
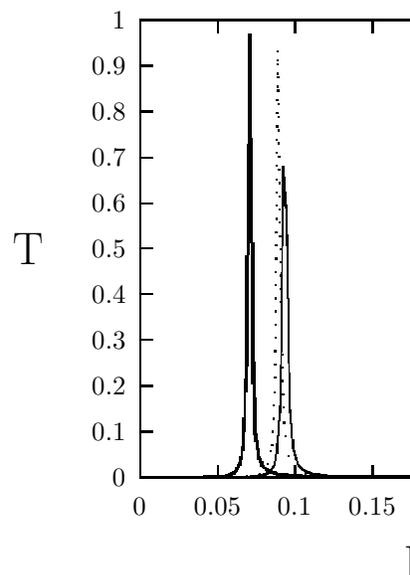
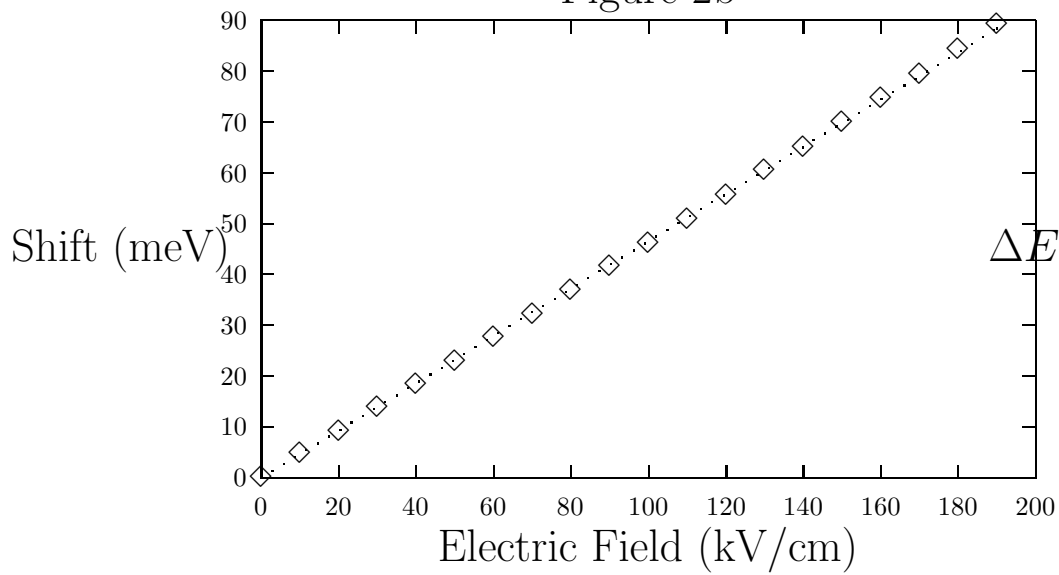
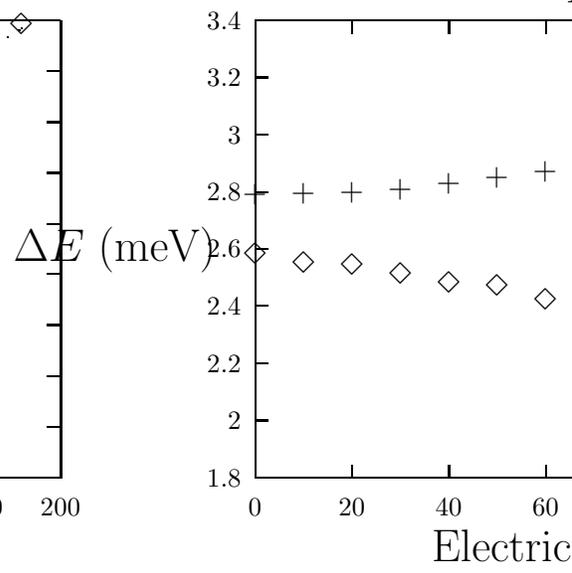
\end{document}